\documentclass[onecolumn,trackchanges]{aastex701}
\makeatletter
\renewcommand{\frontmatter@title@above}{}%
\journalinfo{}
\makeatother
\usepackage[dvipsnames]{xcolor}

\usepackage{float}
\usepackage{subcaption}

\shorttitle{Strong Bars, Strong Inflow}
\shortauthors{M. Magnan et al.}

\begin{document}

\title{Strong Bars, Strong Inflow: The Effect of Bar Strength on Gas Inflow}

\correspondingauthor{Ma\"elle Magnan}
\email{maelle.magnan@mail.utoronto.ca}

\author[orcid=0009-0001-1572-3549]{Ma\"elle Magnan}
\affiliation{David A. Dunlap Department of Astronomy \& Astrophysics,
University of Toronto, 50 St. George St, Toronto, ON M5S 3H4, Canada}
\email{maelle.magnan@mail.utoronto.ca}

\author[orcid=0000-0002-6851-9613]{Tobias G\'eron}
\affiliation{Dunlap Institute for Astronomy \& Astrophysics,
University of Toronto, 50 St. George St, Toronto, ON M5S 3H4, Canada}
\email{tobias.geron@utoronto.ca}

\author[orcid=0000-0002-3887-6433]{Izzy L. Garland}
\affiliation{Department of Theoretical Physics and Astrophysics, Faculty of Science, Masaryk University, Kotl\'{a}\v{r}sk\'{a} 2, Brno, 611 37, Czech Republic}
\email{garland@mail.muni.cz}

\author[orcid=0000-0001-5578-359X]{Chris J. Lintott}
\affiliation{Oxford Astrophysics, Department of Physics, University of Oxford
Denys Wilkinson Building, Keble Road
Oxford, OX1 3RH, UK}
\email{chris.lintott@physics.ox.ac.uk}

\author[orcid=0009-0009-6545-8710]{Jason Shingirai Makechemu}
\affiliation{Department of Physics, Lancaster University, Lancaster, LA1 4YB, UK}
\email{j.makechemu@lancaster.ac.uk}

\author[orcid=0000-0003-1217-4617]{David O'Ryan}
\affiliation{European Space Agency (ESA), European Space Astronomy Centre (ESAC), Camino Bajo del Castillo s/n, 28692, Villanueva de la Cañada, Madrid, Spain}
\email{doryan@cab.inta-csic.es}

\author[orcid=0000-0001-5882-3323]{Brooke D. Simmons}
\affiliation{Department of Physics, Lancaster University, Lancaster, LA1 4YB, UK}
\email{b.simmons@lancaster.ac.uk}

\author[orcid=0000-0001-6417-7196]{Rebecca J. Smethurst}
\affiliation{Oxford Astrophysics, Department of Physics, University of Oxford
Denys Wilkinson Building, Keble Road
Oxford, OX1 3RH, UK}
\email{rebecca.smethurst@physics.ox.ac.uk}

\begin{abstract}
Stellar bars are elongated structures in disk galaxies that can torque and funnel gas inward, influencing galaxy evolution. While strong bars are known to induce rapid inflow, the impact of weaker bars remains less certain. We collected spectroscopic data using the Isaac Newton Telescope to analyze 18 nearby galaxies (strongly barred, weakly barred, and unbarred) drawn from Galaxy Zoo DESI.  We obtained spatial profiles of equivalent width (EW) and ionized gas velocity dispersion by fitting Gaussian profiles to the H$\alpha$ emission line. Strongly barred galaxies exhibit a distinctive three-peaked EW[H$\alpha$] structure, consistent with inward funneling of gas. Weakly barred systems lack this pattern, which suggests limited inflow. Velocity dispersion distributions further distinguish the bar types, with strongly barred galaxies showing significantly higher values than weakly barred and unbarred systems. These results suggest that strong bars drive gas inflow, while weak bars exert a limited dynamical influence.
\end{abstract}






\section{Introduction} 
Secular (non-merger) processes are a crucial part of galaxy evolution, with stellar bars playing a major role (e.g. \citealt{Sellwood}). Stellar bars are common structures, present in roughly half of local disk galaxies \citep{Menendez2007}. By exerting gravitational torques that redistribute angular momentum between stars and gas, bars can drive gas inflow toward the central regions of galaxies.  
This inflow can trigger central starbursts or dynamically heat the gas, in either case suppressing star formation and contributing to quenching \citep{Géron2021}.  Understanding how bar strength regulates these processes remains central to models of secular galaxy evolution. 

In the Galaxy Zoo survey, galaxies are classified as strongly barred, weakly barred, or unbarred \citep{GalaxyZoo}. Strong bars have been found to facilitate the quenching process via rapid bar-driven inflow \citep{Géron2021}. Gas is funneled efficiently to the central regions, depleting the outskirts while increasing central star formation and ultimately exhausting or stabilizing the remaining gas. In contrast, weak bars do not power quenching, and their impact on gas inflow and turbulence remains largely unstudied. 


In this work, we investigate how bar strength affects the inflow and turbulence of ionized gas using H$\alpha$ spectroscopy of nearby galaxies from Galaxy Zoo DESI. By comparing spatial profiles of EW[H$\alpha$] and velocity dispersion across strongly barred, weakly barred, and unbarred systems, we aim to determine whether weak bars can drive measurable gas inflow, and to quantify how bar strength shapes the dynamical and star-forming evolution of disk galaxies.

\begin{figure}[h]
\centering
\includegraphics[width=\linewidth]{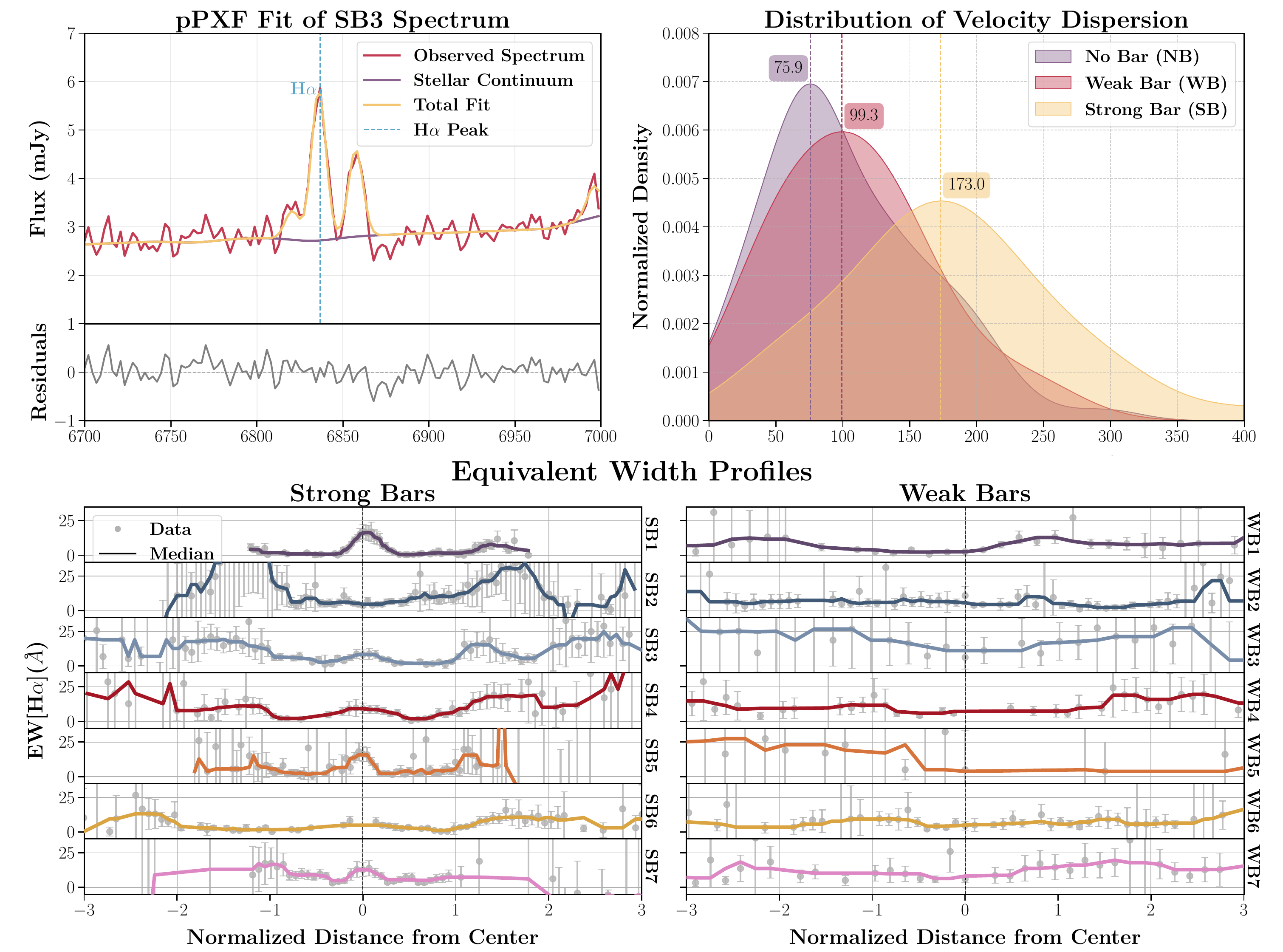}

\caption{\textbf{Top left:} Example \textsc{pPXF} fit to the H$\alpha$ emission line for a strongly barred galaxy (SB3). The upper panel shows the flux-calibrated observed spectrum (maroon), the stellar continuum derived from MILES templates (pink), and the total model (yellow), while the lower panel shows the residuals between the data and the model. \textbf{Top right:} Kernel density estimates of the velocity dispersion distributions for no bar (maroon), weak bar (pink), and strong bar (yellow) galaxies. Densities are normalized by the number of data points in each subsample, and vertical dashed lines indicate the median velocity dispersion for each class. \textbf{Bottom:} Equivalent-width profiles of H$\alpha$ for strongly barred (left) and weakly barred (right) galaxies. Gray points with error bars show the individual EW[H$\alpha$] measurements along the bar as a function of normalized distance from the galaxy center. The colored curves indicate a three-point running median of the measurements. The vertical dashed line marks the bar center ($R=0$).}
\label{fig:plot}
\end{figure}

\section{Methods} 
We used Galaxy Zoo DESI to select 18 galaxies for our study: 7 strongly barred (SB), 7 weakly barred (WB), and 4 non-barred (NB) galaxies. The dataset was limited to galaxies within $0.01<z<0.05$, and an absolute magnitude cutoff of $M_R < -18.96$ to create a volume limited sample of nearby galaxies with resolved bars. We selected targets that were part of the ALFALFA survey, to confirm the presence of interstellar gas \citep{ALFALFA}. To avoid biases introduced by variations in stellar mass, we selected each SB-WB-NB triplet to have similar stellar masses. 

Spectroscopic observations of the galaxies were carried out with the 2.5~m Isaac Newton Telescope, at the Roque de los Muchachos Observatory on La Palma, using the Intermediate Dispersion Spectrograph equipped with the EEV10 detector and R300V grating \citep{INT}. The raw images were reduced using a standard calibration procedure. 

We performed gas emission-line fitting to measure the kinematics of the ionized gas in each galaxy. Among the available optical lines, H$\alpha$ is the brightest, making it an effective tracer of ionized gas and its internal motions. We employed the Penalized Pixel-Fitting ({\textsc{pPXF}}) algorithm to model the emission lines and extract their kinematic parameters \citep{cappellari2023}. The fit included H$\alpha$, and [\ion{N}{2}]~$\lambda6548,\lambda6583$, [\ion{S}{2}]~$\lambda6716,\lambda6731$, whose proximity to H$\alpha$ helps to constrain measurements. Emission lines were modeled using Gaussian templates, while the underlying stellar continuum was fitted using MILES simple stellar population templates \citep{MILES}. The top left panel of Fig. \ref{fig:plot} shows an example of a \textsc{pPXF} fit. We extract the equivalent width and velocity dispersion from the H$\alpha$ fits and propagate errors based on the noise estimates output by \textsc{pPXF}. 


\section{Results \& Discussion}
The H$\alpha$ equivalent width profiles are shown in the bottom panel of Fig. \ref{fig:plot}. Many of the SB profiles exhibit a characteristic three-peaked structure, with a prominent central maximum and two smaller peaks near the bar ends. This pattern suggests gas inflow along the bar: gas accumulates in the central region while residual material is left behind near the bar end. This result is consistent with previous work reporting similar H$\alpha$ profile morphologies \citep{2020MNRAS.495.4158F, Geron2024}. WB galaxies lack this three-peaked structure, providing no clear evidence for bar-driven gas inflow.

The ionized gas velocity dispersion traces turbulence in the galaxy. Bar-driven inflow naturally induces turbulence through shocks and collisions along the bar, thus elevated dispersions are expected for stronger inflow. To compare bar types, we combined individual galaxy measurements within each class into a single population. This approach allowed us to characterize the overall dispersion distribution associated with each bar type. The top right panel of Fig. \ref{fig:plot} shows kernel density estimates (KDEs) of these aggregated distributions. The KDEs reveal that strongly barred galaxies exhibit systematically higher velocity dispersions than weakly barred or unbarred systems. The median velocity dispersion in SB galaxies is $173.0~\mathrm{km\,s^{-1}}$, compared to $99.3~\mathrm{km\,s^{-1}}$ in WB systems and $75.9~\mathrm{km\,s^{-1}}$ in NB galaxies. The higher velocity dispersions observed in SB systems are consistent with bar-driven inflow toward the central regions, whereas the lower dispersions measured in weakly barred galaxies suggest more limited central inflow. The Anderson–Darling $k$-sample test further supports these findings: the comparison between SB and WB galaxies yields a highly significant difference with $p < 0.001$ ($>3.3\sigma$), and the SB versus NB comparison shows an equally strong result with $p < 0.001$ ($>3.3\sigma$). In contrast, the WB versus NB comparison is not statistically significant, with $p > 0.25$ ($<1.15\sigma$).


Overall, these results indicate that SB galaxies are dynamically distinct from both WB and NB systems, while WB galaxies occupy an intermediate state. Strongly barred galaxies combine three-peaked EW[H$\alpha$] profiles with systematically higher ionized gas velocity dispersions, both consistent with enhanced bar-driven inflow and turbulence. Weakly barred systems show flatter EW profiles and lower dispersions, indicating limited inflow.
We will extend this study to a larger sample in the future, using a survey such as MaNGA \citep{Manga}.


\bibliography{bars}{}
\bibliographystyle{aasjournalv7}



\end{document}